\documentstyle[twocolumn,prl,aps]{revtex}
\tighten
\begin{document}
\title{The Glass Transition of Thin Polymer Films: Some Questions, and a Possible
Answer}
\author{Stephan Herminghaus, Karin Jacobs, Ralf Seemann}
\address{Dept. of Applied Physics, University of Ulm, D-89069 Ulm,
Germany}
\date{\today}
\maketitle

\begin{abstract}

A simple and predictive model is put forward explaining the
experimentally observed substantial shift of the glass transition
temperature, $T_g$, of sufficiently thin polymer films. It focuses
on the limit of small molecular weight, where geometrical `finite
size' effects on the chain conformation can be ruled out. The
model is based on the idea that the polymer freezes due to memory
effects in the viscoelastic eigenmodes of the film, which are
affected by the proximity of the boundaries. The elastic modulus
of the polymer at the glass transition turns out to be the only
fitting parameter. Quantitative agreement is obtained with our
experimental results on short chain polystyrene ($M_W$ = 2
kg/mol), as well as with earlier results obtained with larger
molecules. Furthermore, the model naturally accounts for the weak
dependence of the shift of $T_g$ upon the molecular weight. It
furthermore explains why supported films must be thinner than free
standing ones to yield the same shift, and why the latter depends
upon the chemical properties of the substrate. Generalizations for
arbitrary experimental geometries are straightforward.
\end{abstract}

\section*{introduction}

To explain the experimental observation that thin polymer films
melt at temperatures well below the glass transition temperature
of the bulk polymer, $T_g^{0}$ \cite{KJ1,MFB,DVF,DVDG}, is one of
the major current challenges in the theory of homopolymers. At
first glance, one might expect such a behavior, due to the impact
of the finite size geometry of a thin film upon objects as large
as polymer molecules. In fact, for molecular weights larger than
about $M_W$ = 300 kg/mol, geometry effects have been recently
shown to play a dominant role, resulting in a linear dependence of
$T_g$ upon the film thickness, $h$ \cite{DVDG,DeGennes,DeGennes2}.
For smaller molecules, however, the problem is conceptually more
intricate, since a noticeable reduction of $T_g$ can be observed
at film thicknesses orders of magnitude larger than the radius of
gyration of the molecules. Furthermore, the shift $T_g^{0}-T_g(h)$
becomes strongly nonlinear and largely independent of molecular
weight \cite{DVF,Forrest}, indicating that this regime must be
governed by a different mechanism. A fundamental understanding of
this effect would be of great interest not only for polymer
physics, but potentially also elucidate the physics of the glass
transition in a larger class of systems, since it comes into play
at small molecular weight.

Many attempts have been undertaken to explain these observations,
mostly by considering microscopic models of the inner structure of
the films. D. Long and F. Lequeux have envisaged the freezing of
the film as a percolation of rigid domains \cite{Moukarzel},
mediated by thermal fluctuations \cite{LongLequeux}. Other models
assume a layered structure of the film, with a particularly mobile
region close to the free surface of the film \cite{KJ1}. Within
this framework, J. Forrest and J. Mattsson \cite{Forrest} have
recently been able to achieve quite impressive accordance with the
experimental data \cite{MFB,Forrest}. Their model makes use of the
so-called cooperativity length, $\xi(T)$, which plays a mayor role
in a whole class of theoretical concepts of the glass transition.
The only drawback is that there is yet no well established theory
of $\xi(T)$. Furthermore, as a consequence of the two-layer
structure of the film inherent in the model, it is not completely
clear why there should not be two glass transitions, rather than a
single one shifted in temperature.

As a complementary approach, computer simulations of polymer films
with free surfaces have recently been carried out, and most of the
experimental findings were reproduced qualitatively \cite{Torres}.
However, the polymer chains in these simulations were only 16
monomer units long, very much shorter than those used in the
experiments. Before attempting to explain the effect
theoretically, it is therefore worthwile to study experimentally
the behavior of polymers with short chain length, both to ease
comparison with simulation and to explore the range of validity of
the apparent independence of the shift of $T_g$ on molecular
weight.

Before going into the details of our study, let us take a break
and summarize the main questions to be answered.
\begin{enumerate}
\item{What is the principal mechanism responsible for the reduction of $T_g$
in thin films of low molecular weight ($M_W<$300kg/mol) polymers?}
\item{Down to how small molecular weight is this mechanism valid?}
\item{Why is the effect stronger in free standing films than in
supported ones \cite{DVF}?}
\item{Why is there no significant dependence of this effect on molecular
weight \cite{DVF}?}
\item{Why does the effect depend upon the chemical composition of
the substrate for supported films
\cite{KJincrease,vanZanten,KJCincrease}?}
\item{Why is there sometimes an increase of $T_g$ in thin films,
instead of a reduction \cite{KJincrease,vanZanten,KJCincrease}?}
\end{enumerate}
We will try in the present paper to give answers to these
questions, or at least show in which direction answers
might be found, on the basis of a novel, quite simple model which rests
mainly on the viscoelastic eigenmodes of the films.
We restrict the discussion to the case of
polystyrene (PS), since this is the most thoroughly
studied polymer in this context. Furthermore, it is particularly well suited for
comparison with theoretical models, since (atactic) polystyrene does not show
any propensity to crystallization. In other polymers, which might crystallize
at least in part of the film, modelling would be exceedingly difficult,
and the main mechanism could well be obscured in experiments.

\section*{Experiment}

Let us first explore the range of validity of the aforementioned
effect of reduction of $T_g$, as to the molecular weight of the
polymer. We have investigated the glass transition in thin
supported films of PS with a molecular weight as small as 2 kg/mol
($\approx 20$ monomer units). It was purchased from Polymer Labs
(UK) with a polydispersity index $M_W/M_N$ = 1.05, the radius of
gyration is 1.3 nm. Effects from the molecular geometry are thus
expected only for films of few nanometers thickness. The films
were spin cast from toluene solution onto silicon wafers (Silchem
GmbH, Freiberg/Germany), which were previously cleaned by
ultrasonication in acetone, ethanol, and toluene, subsequently.
Residual organics were removed with a 1:1 mixture of $H_2SO_4$
with $H_2O_2$, and the substrates were thoroughly rinsed with hot
millipore water afterwards. Films were investigated with
thicknesses ranging from 4 to 160 nm. The roughness of the free
surface of the films was less than 0.2 nm, as revealed by scanning
force microscopy (SFM).

The glass transition temperature was determined in two different
ways, depending on film thickness. The standard procedure of
monitoring the thermal expansion of the film via ellipsometry, as
introduced by Keddy and Jones \cite{KJ1}, was used for film
thicknesses down to 9.6~nm. Fig.~\ref{TypicalRun}a shows a typical
run. The data were reproducible, irrespective of being taken
during heating or cooling. Typical heating or cooling rates were
2~$K/min$. We plotted the changes in refractive index and
thickness of the film in fig.~\ref{TypicalRun}b. The solid line
represents the Clausius Mosotti relation. Obviously, there is good
agreement, suggesting the absence of any loss or degradation of
material. This was found invariably for all samples.

In fig.~\ref{ThermalExpansion} we plotted the thermal expansion
coefficients found above and below $T_g(h)$, which coresponds to
the kink in fig.~\ref{TypicalRun}a. At large film thickness, $T_g$
obtained in this way approached $327 \pm 1$~K, which is consistent
with the temperature at which macroscopic melting is observed in
the bulk ($T_g^{0}$) for PS with this chain length. The shaded
areas represent published data of the expansion coefficients
\cite{ExpansionCoefficient}. Good agreement is found both above
and below $T_g$. Small deviations at very small film thickness, as
might be concluded from our data, have been reported before
\cite{KJ1}.

For films thinner than 9 nm, we determined the melting behaviour
of the films by observing the buildup of amplified thermal
fluctuations (spinodal dewetting) \cite{Karim,OurPRL}. These
processes were monitored by SFM with {\it in-situ} heating. In
order to speed up the experiments to a feasible time scale,
dewetting was observed at temperatures close to $T_g^{0}$. As an
example, we show in fig.~\ref{SpinDewett} the temporal evolution
of the Fourier transform (spatial power spectrum) of the surface
topography. The hallmark feature is the clear peak representing
the fastest growing mode. The inset shows the peak intensity as a
function of time on logarithmic scale. Exponential growth is
clearly observed up to a rather well defined time at which
coalescence of holes sets in.

From the slope of the straight line in the inset, and the known
effective interface potential \cite{OurPRL}, the viscosity $\eta$
of the polymer can be determined. Measuring the viscosity as a
function of temperature, we found that this obeyed a Vogel-Fulcher
law, with the Vogel-Fulcher temperature shifted by a certain
amount $\Delta T_{VF}$, which depended on the film thickness. We
identified $\Delta T_{VF}(h)$ with the shift in glass temperature.
In this way, the apparent glass transition temperature of
particularly thin films was inferred from the spinodal dewetting
experiments by setting $T_g(h)=T_g^0-\Delta T_{VF}(h)$.

Our experimental results are shown in fig.~\ref{Tgresult} as the
full symbols. The circles represent the thermal expansion
measurements, the squares were obtained from the spinodal
dewetting experiments. As one can clearly see, the glass
transition temperature is substantially reduced for all films
thinner than about 50 nm. The solid line represents the function
\begin{equation}
T_g = T_g^0(1+h_0/h)^{-1}
\label{KimFormel}
\end{equation}
This form has been shown before to account well for the data
obtained by others for larger molecular weight films, if $h_0 =
0.68$ nm was assumed for PS \cite{Kim}. Within experimental
scattering, our data exhibit indeed the same dependence of
$T_g(h)$ in the full range of film thickness explored. It is
remarkable that the data from both the thermal expansion and
spinodal dewetting measurments are well fitted by the same curve.
This tentatively corroborates the procedure of obtaining $T_g$ for
very thin films (squares) as discussed above.

Although our polymer chains are roughly by a factor of 50 shorter
than those investigated before, we obtain $h_0 = 0.82$ from the
fit, which is quite close to the above value. This confirms the
weak dependence (if there is a significant one at all) of the
reduction of the glass transition temperature on the molecular
weight of the polymer, down to a molecular weight as small as 2
kg/mol.

We can thus state that the reduction of $T_g$ in thin films, as
described phenomenologically by eq.~(\ref{KimFormel}), is observed
in a huge range of molecular weight, from molecules as small as 2
kg/mol up to a few hundred kg/mol. This is in accordance with
molecular dynamics simulations \cite{Torres}, and rules out mere
finite size effects on the individual coils as the main cause of
the reduction of $T_g$ in this regime. Note that the radius of
gyration of our polymer is only 1.3 nm, while $T_g$ is
significantly reduced at a film thickness of 50 nm already. The
fitting parameter used in eq.~(\ref{KimFormel}), $h_0$, changes
only by about 20 \% in this range, confirming that the dominant
mechanism which is responsible for this effect cannot depend
strongly on the molecular weight of the polymer.

\section*{Eigenmode spectrum of the films}

Since it is clear that the behaviour displayed in
fig.~\ref{Tgresult} can in no way be attributed to the geometrical
impact of the finite film thickness upon the microscopic
conformation of the individual chains, we present here an approach
to the problem which intentionally makes as few reference as
possible to the molecular structure of the film. The latter is
accounted for merely by the strain in the polymer, i.e., the
deviation of the {\it local average} gyration ellipsoids of the
molecules from a sphere. By `{\it local}' we mean a volume much
larger than the volume of the backbone of a single molecule, but
with a lateral dimension much smaller than the film thickness.

Such a deviation from the equilibrium conformation (i.e., from a
Gaussian coil, if self-avoiding is neglected), which may be viewed
as an entropy fluctuation, can decay either by self diffusion of
the individual molecules, or by some center-of-mass rearrangement
(i.e., flow) of the melt. It is clear that close to a surface,
such rearrangements are much easier to accomplish due to coupling
of the capillary waves on the free surface to the bulk flow of the
polymer. This coupling is effective down to a depth comparable to
the wavelength of the capillary modes, which may be {\it much
larger than the coil size of the molecules}. Thus we are provided
with a mechanism which affects the motion of the polymer
molecules, and naturally can act over distances which are large as
compared to molecular dimensions. In order to explore the possible
relevance of this mechanism for the reduction of $T_g$ in thin
films, we have to consider the eigenmode spectrum of the
(viscoelastic) polymer film. This can be discussed with all
possible boundary conditions at the substrate, such that films
with strong slip along the substrate, grafted films, or free
standing films, may as well be treated within the same framework.

The spectrum of a viscoelastic thin film can be
obtained in a straightforward manner by
combining standard theory of elasticity \cite{LL1} and
hydrodynamics \cite{LL2} in the limit of small Reynolds number
(Stokes dynamics). The equation of motion reads
\begin{equation}
\lbrace \partial_t + \omega_0 + \frac{E}{\eta} \rbrace \nabla^2
{\bf \phi} = \frac{\nabla p}{\eta}
\label{EoM}
\end{equation}
where $E$ is Young's modulus, $\eta$ is the viscosity, and
$\omega_0$ is the Rouse rate of relaxation of the individual
chains into their equilibrium configuration \cite{strobl,Kimmich}.
${\bf \phi}$ is a vector field related to the strain tensor, ${\bf
S}$. For the sake of clarity, we restrict our discussion to a
simple model, taking into account only the single intrinsic
relaxation rate $\omega_0$, as opposed to more general treatments
\cite{SafranKlein}. Deviations of the real polymer from this
simple behaviour will be introduced {\it a posteriori} farther
below.

If we restrict the discussion to one lateral ($x$) and one normal
($z$) coordinate, ${\bf \phi} = (\phi_x ,\phi_z)$ is defined via
\begin{equation}
{\bf S} = \left(\begin{array}{cc} \partial_x \phi_x & \frac{1}{2}
(\partial_x \phi_z + \partial_z \phi_x) \\
\frac{1}{2}(\partial_x\phi_z + \partial_z\phi_x) &
\partial_z\phi_z
\end{array}\right) \label{strain}
\end{equation}
Finally, $p$ is the pressure field.

For harmonic excursions of the free surface, $\zeta(x) = \zeta_0
\exp\{iqx-\omega t\}$, eq.(\ref{EoM}) has solutions
\begin{eqnarray}
\phi_x = [1+(h+q^{-1})\alpha(q)]\cosh qz - q^{-1}\alpha(q)\sinh qz
\nonumber \\ \phi_z = [1+h\alpha(q)]\sinh qz - z\alpha(q)\cosh qz
\label{solutionfree}
\end{eqnarray}
where for the function $\alpha(q)$, we find
\begin{equation}
\alpha(q) = \left(\frac{q}{2}\right) \frac{e^{2qh}-1}{e^{2qh}-1+qh}
\label{alphafree}
\end{equation}
for free standing films (symmetric modes) as well as for supported
films with full slippage (zero friction). For supported films with
some friction at the substrate, the expressions are of similar
form, but considerably more complicated, and will not be discussed
here. At the free surface, we used the standard boundary condition
of zero tangential stress, and $p = -\sigma\partial_{xx}\zeta$,
where $\sigma$ is the surface tension of the polymer. Note that
for free standing films, $h$ is defined as {\it half} the film
thickness.

For the relaxation rates of the modes, we get
\begin{eqnarray}
2\omega = (\omega_0 + \frac{E}{\eta} + \frac{\sigma q^2}{2 \eta
\alpha(q)}) \nonumber \\
\pm \sqrt{\left(\omega_0 + \frac{E}{\eta} + \frac{\sigma q^2}{2
\eta \alpha(q)}\right)^2 + \omega_0\frac{2 \sigma q^2}{\eta
\alpha(q)}} \label{dispersion}
\end{eqnarray}
Since it is only the `fast' modes which contribute appreciably to
the reduction of the glass transition temperature (see below), we
consider only the upper branch of eq.~(\ref{dispersion}).
Observing that $\omega_0 << \frac{E}{\eta}$, this is given by
\begin{equation}
\omega = \omega_0 + \frac{E}{\eta} + \frac{\sigma q^2}{2 \eta
\alpha(q)},
\label{dispersionbranch}
\end{equation}
as a very good approximation. The dependence of the wave number,
$q$, stems solely form the coupling to the capillary waves on the
free film surface. The first major assumption of our model is that
$T_g(h)$ is determined by the spectrum of the viscoelastic
eigenmodes of the film as given by eq.~(\ref{dispersionbranch}).

\section*{The freezing mechanism: memory effects}

The second major assumption is that the physical cause for the
melting or freezing of the film, respectively, are {\it memory
effects} in the polymer material. These are of course not included
in the linear theory discussed above, and may be formulated in a
generic way by means of a suitable memory kernel, as used in a
class of theoretical models of the glass transition, called mode
coupling models \cite{Goetze1,Goetze2}. Within this framework,
memory effects are taken into account by inserting a convolution
integral with the memory kernel $m\{\phi(t)\} = a_1 \phi + a_2
\phi^2 + a_3 \phi^3 + ...$ in the otherwise linear differential
equation of motion of the modes considered
\protect\cite{footnotestrobl}. In these models, $\phi$ usually
describes density fluctuations. In contrast, we consider the
material to be essentially incompressible, and $\bf\phi$ denotes
here the strain in the polymer material, as defined above. It thus
describes the local state of the material, similar to what the
density does for simple glass forming liquids
\cite{Goetze1,Goetze2,Zaccarelli}.

In our case, the equation of motion (2) is of first order in time,
hence we have
\begin{equation}
{\bf\phi}^{\prime} + \omega(q) \phi + \int_0^t
m\{{\bf\phi}(\tau)\}{\bf\phi}^{\prime}(t-\tau)d\tau\ = 0
\label{MCTrelax}
\end{equation}
as the mode coupling equation. This type of equations has been
thoroughly analysed \cite{Goetze1,Goetze2,Leutheusser} in relation
to the {\it microscopic} physics of the glass transition (to which
we do not refer here), as well as to large scale degrees of
freedom \cite{Lequeux}. For density fluctuations in glass forming
simple liquids, it was found that the coefficients $a_i$ of the
memory kernel vary concurrently with temperature, and that upon
crossing a certain border in the space spanned by the $a_i$, the
system freezes into a nonergodic state \cite{Goetze2}. The
existence of such a freezing transition has been found to be
largely independent of the precise form of the memory kernel. In
fact, most of the features of a glass transition may be well
represented in what has become known as schematic models, which
are simple mode coupling equations not referring in their memory
kernels to the microscopic physics of the system under study.

It is now worthwhile to contemplate on possible memory effects in
a polymer melt, as to their scaling with temperature. Let us first
consider a polymer molecule in equilibrium, forming a more or less
Gaussian coil. If this is elongated by straining the polymer melt,
to what extent will it memorize this process after the strain is
released? In the strained state, the molecule will relax to some
extent. However, this relaxation will not proceed homogeneously
along the molecule, since the activation energy, $U$, for local
rearrangements will depend upon the local topological environment.

In order to discuss the relaxation behaviour of the local
molecular geometry, let us define a local geometric exponent of
the coil, $\mu$, by the relation
\begin{equation}
<\Delta\mid{\bf r}\mid> = (\Delta s)^{\mu}
\label{geometricexponent}
\end{equation}
where $<\Delta\mid{\bf r}\mid>$ is the typical distance travelled
in space upon moving alongside the polymer chain by a distance
$\Delta s$. In completely stretched parts of the chain, $\mu = 1$,
whereas in equilibrated domains, $\mu = \mu_{equilib.} \approx
0.5$. The variation $\Delta U$ of $U(s)$ determines the variation
of the relaxation rate of $\mu$ towards $\mu_{equilib.}$. When the
strain is released, remnants of this variation will remain, and
thus represent a memory of the strain. This holds as long as the
strain/release process is fast as compared to the equilibration
time, $\omega_0^{-1}$. For the modes to be considered here, this
is well fulfilled. When $\Delta U << kT$, as can be safely
assumed, these remnant variations in $\mu(s)$ scale as $\Delta
U/kT$, thus we can conclude that memory effects in the polymer
scale as $1/T$. As a direct consequence, we can replace the memory
kernel $m\{ {\bf \phi (t)}\}$ of eq.~(\ref{MCTrelax}), the
coefficients $a_i$ of which are temperature dependent, with
$M\{\phi\}/T$, where $M$ is now independent of temperature.

By proper normalizaton of time, eq.~(\ref{MCTrelax}) can now be
rewritten as
\begin{equation}
\phi^{\prime} + \phi + \frac{1}{T\omega(q)}\int_0^t
M\{\phi(\tau)\}\phi^{\prime}(t-\tau)d\tau\ = 0
\label{MCTnormalized}
\end{equation}
From this equation, we see directly that the modes with the
largest relaxation rate freeze at the lowest temperature.
Furthermore, the precise form of $M$, which might be derived from
a detailed analysis of the non-equilibrium dynamics of the polymer
molecules, does not need to be considered here any further. It is
only required that it belongs to the class of kernels which yield
a freezing transition for ${\bf\phi}$ at all.

Let us now turn back to the eigenmodes of the film. The relaxation
rates, as given by eq.~(\ref{dispersionbranch}), are monotonously
increasing with $q$. However, modes with $q$ much larger than the
inverse film thickness, $h^{-1}$, do not penetrate appreciably
into the film, such that only a small fraction of the material
takes part in these modes. Hence we are led to considering chiefly
the modes with $q \approx h^{-1}$, since these are the highest
frequency modes comprising all of the film material.

Setting thus $T_g\omega(h^{-1})=$\ {\it const.} as suggested by
eq.~(\ref{MCTnormalized}), we directly arrive at a simple formula
for the glass transition temperature:
\begin{equation}
T_g(D) = T^0_g\left(1+\frac{1.16 \
\sigma}{h(E+\eta\omega_0)}\right)^{-1} \label{result}
\end{equation}
This is precisely the form of eq.~(\ref{KimFormel}), and
fig.~\ref{Tgresult} shows that it describes our data very well.

\section*{Discussion}

The quantity $\eta \omega_0$ is on the order of a few kPa and can
in general be safely neglected against $E$. The characteristic
length scale which appears here, and which was called $h_0$ in
eq.~(\ref{KimFormel}), is thus the ratio $\sigma/E$, with the
surface tension $\sigma = 31$ mN/m for PS. It follows that the
elastic modulus determining the dynamics of the relevant modes,
$E$, is {\it the only physical fitting parameter} in the model.
From $h_0 = 0.82$ nm, as obtained from the fit of our model, we
find for the elastic modulus $E \approx 44$~MPa, which is, on
logarithmic scale, right in between the modulus of the frozen
material (a few GPa) and the modulus just above $T_g$ (about 300
kPa). On the basis of the model discussed above, let us now try to
formulate answers to the questions asked in the introduction, one
after the other.

1. \ What is the main mechanism responsible for the reduction of
$T_g$ in thin films of low molecular weight ($M_W<$300kg/mol)
polymers? The physical picture which emerges from our model is
that as the temperature is increased, melting proceeds as the
fastest mode involving all of the film material escapes from its
frozen state and fluctuates. This proceeds with the help of the
capillary waves on the free film surface, which ease entropy
fluctuations in the polymer melt. The concomitantly increased
motion of polymer chains reduces the effective viscosity also for
modes with smaller $q$, which, as a consequence, are sped up
($\eta$ appears in the denominator in
eq.~(\ref{dispersionbranch})!) and melt in turn. In this way, the
film finally melts at all length scales. It is illustrative to
note that the softness of the spinodal modes, as used for the
determination of $T_g$ at small film thickness, does not
contradict our model: these are modes with $q<<1/h$, and are thus
not the fastest modes, as considered in the melting mechanism.

2. \ Down to how small molecular weight is this mechanism valid?
We found that in agreement with simulation results \cite{Torres},
the effect is present even for molecules as short as about 20
monomer units. This is well below the limit of entanglement. It
thus spans a wide range in molecular weight, over more than two
orders of magnitude. It is precluded, or obscured, by other
effects at molecular weights in excess of about 300 kg/mol
\cite{DVF}.

3. \ Why is there no significant dependence of the effect on
molecular weight? The physical property which solely determines
the thin film behaviour according to our model is the elastic
modulus, $E$. As it is well known, $E$ displays no marked
dependence on the molecular weight \cite{strobl,stoll}, such that
on the basis of our model, $T_g(h)$ is expected as well to be
largely independent of molecular weight. To be precise, $E$ is
slightly less for smaller molecular weight than for larger
\cite{ExpansionCoefficient,stoll}, such that according to
eq.~(\ref{result}), $h_0$ should be larger for smaller molecular
weight. In fact, we obtained $h_0=0.82$ nm for our very short
molecules, whereas for larger molecules, $h_0=0.68$ nm was found
\cite{Kim}. The observation that the relevant length scale, $h_0$,
decreases with increasing molecular weight is another evidence
against geometrical effects on individual chains to be relevant in
the regime discussed here. The exact physical significance of the
somewhat arbitrary `choice' of $E$, which may be viewed as $E$
{\it at} $T_g$, is to be investigated in further studies.

4. \ Why is the effect stronger in free standing films than in
supported ones? We have mentioned that the eigenmodes of supported
films of thickness $h$ with no friction are identical to those
(the symmetric ones) of free standing films of thickness $2h$.
Thus our model predicts that the effect in a supported film is
just as large as in a free standing film of twice the thickness.
This is indeed in accordance with experimental observation
\cite{DVF}.

5. \ Why does the effect depend upon the chemical composition of
the substrate for supported films? It is clear that the amount of
friction of the film material at the substrate changes the
eigenmode spectrum of the film, such that a dependence on the
chemical composition of the substrate is indeed expected on the
basis of our model. For a quantitative comparison with
experimental data, however, the friction coefficients would have
to be determined for the systems investigated. It should be well
noted that this also induces some uncertainty in the numerical
value of $E$ derived from our fit (cf. fig.~\ref{Tgresult}), since
we have not characterized the friction of the films on the
substrate.

6. \ Why is there sometimes an increase of $T_g$ in thin films
\cite{KJincrease,vanZanten,KJCincrease}? We can try to give a
rough idea to what extent effects like this may be accounted for
within our model, by including the interaction of the free surface
with the substrate via long range forces. This is described most
conveniently by replacing $\sigma q^2$ by $\sigma q^2 +
\frac{d^2V}{dh^2}$, where $V(h)$ is the effective interface
potential of the film due to long range forces
\cite{Dietrich,Schick}. For unretarded van der Waals forces,
$V(h)=-A/12\pi h^2$, where $A$ is the Hamaker constant. If $A>0$,
$\frac{d^2V}{dh^2}$ is negative, such that the van der Waals
forces tend to destabilize film, giving rise to spinodal dewetting
for sufficiently thin films. In any case, a positive $A$ will
reduce the relaxation rate of all of the modes, including the one
at $q=1/h$, and thereby tend to increase the glass transition
temperature. Quantitatively, this effect is obtained by
accordingly replacing $\sigma$ in eq.~(\ref{result}) by $\sigma -
\frac{A}{2\pi h^2}$. As it turns out, $T_g$ attains a minimum at
$h_{min}=3\sqrt{A/2\pi\sigma}$, and increases sharply for smaller
thickness. For $A = 2.2 \times 10^{-20}$ J, representing PS on
silicon oxide \cite{GainingControl}, this is at
$h_{min}\approx$~1~nm. An increase of $T_g$ above $T_g^0$ at a
film thickness significantly larger than 1~nm can be explained
only with an unphysically large Hamaker constant.

However, one should anyway be cautious with trying to interpret an
increase of $T_g$ in thin films in the framework of our model if
working with polymers other than PS
\cite{Torres,KJincrease,vanZanten,KJCincrease}. It is not at all
clear what impact a possible crystallization (which is not present
in atactic PS) can have on $T_g(h)$. Consequently, possible model
systems for studying the mechnaism discussed in the present paper
have to be carefully chosen to ensure the absence of any
pronounced inner texture of the film. Our model can be appropriate
only for `structureless' films, and might break down for films
with a pronounced layered texture like, e.g., a Langmuir-Blodgett
film \cite{LBpolymer}, or for other polymers which may crystallize
at least in some part of the film.

\section*{Outlook}

It is tempting to apply the view developed here also to the
surface of a bulk polymer sample, for which the spectrum is
obtained setting $\alpha = q/2$. As it is readily seen, one should
expect surface melting at the polymer surface, down to a thickness
of $h_{sm} = h_0 T_g^{0}(T_g^{0}-T)^{-1}$. This is very much along
the lines of first ideas which had been put forward to explain the
observed reduction of $T_g$ in thin films \cite{KJ1}. To the best
of our knowledge, there is yet no conclusive evidence in favor or
in disfavor of polymer surface melting.

To confirm the theoretical concept put forward here, an {\it
ab-initio} derivation of the memory kernel, $m\{{\bf \phi}\}$,
from the statistics and microscopic transport properties of the
individual molecule would be desirable. Furthermore, it is
important to elaborate on the significance of Young's elastic
modulus, $E$, at the glass transition. This was determined
experimentally using our model to be $E = 44$~MPa, but it is not
yet clear what physical principle distinguishes this value. A
complete theory of the glass transition on the basis of the
proposed model would be rather cumbersome, since it involves two
strongly nonlinear mechanisms mutually affecting each other:
freezing by the mode coupling mechanism, and non-Newtonian effects
on the viscosity. Our model furthermore suggests that the glass
transition in a homopolymer may be viewed as a mode-coupling-model
freezing of its viscoelastic bulk modes. This needs not to
contradict more classical views of the glass transition in
homopolymers, but might serve as an interesting, and potentially
useful, alternative approach.

\acknowledgements The authors owe many very helpful hints to J. A.
Forrest and K. Dalnoki-Veress. We are furthermore indebted to D.
Johannsmann, J. Baschnagel and R. Blossey for stimulating
discussions. Funding from the Deutsche Forschungsgemeinschaft
within the Priority Program `Wetting and Structure Formation at
Interfaces' is gratefully acknowledged.

\begin{figure}
\caption{a) Ellipsometric dilation measurement of a thin PS film.
Typical heating/cooling rates were 2 K/min. b) The change in film
thickness vs. the change in the optical dielectric constant, as
measured by ellipsometry for various temperatures during a
heating/cooling cycle. The good agreement obtained with the
Clausius-Mosotti relation represented by the solid line
demonstrates the absence of significant loss or degradation of
material.} \label{TypicalRun}
\end{figure}

\begin{figure}
\caption{Thermal expansion coefficients obtained above and below
$T_g$. The shaded areas represent values taken from the literature
\protect\cite{ExpansionCoefficient}.} \label{ThermalExpansion}
\end{figure}

\begin{figure}
\caption{Temporal evolution of the power spectrum of spinodal
dewetting undulations on a PS film exhibiting unstable capillary
waves at its free surface. Curves of equidistant times are
superimposed to show the gradual increase of the main peak. The
inset shows the exponential growth of the latter, from which the
viscosity of the film can be inferred.} \label{SpinDewett}
\end{figure}

\begin{figure}
\caption{The glass transition of thin films of 2 kg/mol
polystyrene, as determined from thermal expansion (circles) and
from the growth of spinodal waves (squares). The solid curve
represents our model, which has the elastic modulus governing the
dominant modes as its only fitting parameter. It furthermore
corresponds to what was found for larger molecular weight PS
before \protect\cite{Kim}. Top: linear scale. Bottom: logarithmic
scale, showing more details at small film thickness.}
\label{Tgresult}
\end{figure}

\end{document}